# Dimensions: re-discovering the ecosystem of scientific information

**Enrique Orduña-Malea** and **Emilio Delgado López-Cózar**

**Abstract:** The overarching aim of this work is to provide a detailed description of the free version of *Dimensions* (new bibliographic database produced by *Digital Science* and launched in January 2018). To do this, the work is divided into two differentiated blocks. First, its characteristics, operation and features are described, focusing on its main strengths and weaknesses. Secondly, an analysis of its coverage is carried out (comparing it *Scopus* and *Google Scholar*) in order to determine whether the bibliometric indicators offered by *Dimensions* have an order of magnitude significant enough to be used. To this end, an analysis is carried out at three levels: journals (sample of 20 publications in 'Library & Information Science'), documents (276 articles published by the *Journal of informetrics* between 2013 and 2015) and authors (28 people awarded with the *Derek de Solla Price* prize). Preliminary results indicate that *Dimensions* has coverage of the recent literature superior to *Scopus* although inferior to *Google Scholar*. With regard to the number of citations received, *Dimensions* offers slightly lower figures than *Scopus*. Despite this, the number of citations in *Dimensions* exhibits a strong correlation with *Scopus* and somewhat less (although still significant) with *Google Scholar*. For this reason, it is concluded that *Dimensions* is an alternative for carrying out citation studies, being able to rival *Scopus* (greater coverage and free of charge) and with *Google Scholar* (greater functionalities for the treatment and data export).

## 1. Introduction

In the vast majority of scientific disciplines based –or largely dependent- on quantitative analysis methods, the measurement tools turn out to be real bottlenecks. The advances in the development of these tools and their capacities and features have a definitive influence on the accuracy of the data obtained, on the correct interpretation of them, as well as on the orientation towards promising lines of research, facilitating some and preventing others. The tool is interposed between the researcher and the observed reality. For example, the development and evolution of lenses (applied to telescopes, microscopes, or contact lens) gives good faith of the advances in Astrophysics, Medicine, Molecular Chemistry and Optics, among other fields (**Delgado-López-Cózar** et al., 2017).

Scientometrics and Bibliometrics (and all the derived "-metrics") are by no means exception. As fundamentally quantitative disciplines (which does not mean that they cannot address more qualitative studies, fortunately) heavily depend on measuring instruments, among which, above all, the bibliographic databases providing information on the citations received by the indexed bibliographic records stand out. It is not surprising therefore that when Information and Communication Technologies made possible the development and implementation of this type of products, the discipline took a giant leap, transforming itself at the epistemological level. The appearance of *Science citation index* (SCI) and *Social science citation index* (SSCI) allowed for the first time the realization of studies that had been impossible and unthinkable to date, contributing to be acquainted with aspects that until then had remained invisible to the eyes of researchers.

The appearance in 2004 of both *Scopus* and *Google Scholar* represents a turning point in the story. However, while *Scopus* gave bring forth evolution (extended coverage, new journal's topic classification, new journal indicators, innovative visualization techniques, and lately an integration of altmetrics) *Google Scholar* implied a revolution. This database claims to identify automatically all online academic material (putting the focus on the article instead of the source publication) bypassing all the peer-reviewed journals' quality filters. Moreover, it integrated the product intuitively with a search engine that used the *Page rank*'s philosophy and its relevance to locate and discover academic material with simplicity and speed (and free of charge), entering the bibliographic databases in the big data universe (**Orduña-Malea** et al., 2016).

The story presents a new milestone on January 15, 2018, the day on which the *Dimensions* platform was officially launched (**Schonfeld**, 2018). This database is endorsed by *Digital science* (a technology company founded in 2009 that funds innovative business oriented to make the different parts of the scientific process more open, efficient and effective), through six companies in its portfolio (*ReadCube*, *Altmetric*, *Figshare*, *Symplectic*, *ÜberResearch* and *Digital science consultancy*).
*https://www.dimensions.ai*
*https://www.digital-science.com*
*https://www.readcube.com*
*http://www.altmetric.com*
*http://figshare.com*
*http://symplectic.co.uk*
*http://www.uberresearch.com*
*https://www.digital-science.com/products/consultancy*

Since the wide amount of data and the existence of ecosystems of research are increasingly diverse, *Dimensions* rises with the purpose of becoming "a modern and innovative infrastructure and linked research data tool" whose purpose is to tear down existing data silos using new technologies. For this, they have started from the underlying technology to a preliminary version launched by *ÜberResearch* in 2014, although according to the parent company itself, the six companies decided to embark on this project together in 2011.

The original idea of the database is to facilitate the identification of experts and leaders in the different scientific domains and, therefore, to favour and stimulate academic networking and partnership (**McShea**, 2018). Similarly, it aims to provide scientists and groups with the design of technological surveillance systems to keep abreast of the latest advances in their various fields. To that end, the database aims to show and connect information from the first signs of academic activity (funded projects) to the last stages (publication in journals and dissemination in social networks), going through a wide variety of document types.

At the time of its launch, *Dimensions* is made up of 128 million documents (among others, 89 million articles, 34 million patents, 380 thousand clinical trials and 320 thousand policy documents) apart from information on funding (3.7 million of awarded grants), and approximately 4 billion connections between them. Additionally, these data are enriched with impact information, both in terms of citations received (connections among cited / citing documents, available for 50 million records) and altmetrics (available for 9 million documents approximately), academic profiles (20 million profiles), Global research identifier database (GRID) geotagging, as well as a classification of subject areas based on machine learning techniques (**Bode** et al., 2018).

The database is offered in three different versions, a free version (*Dimensions*) and two paid versions (*Dimensions plus* and *Dimensions analytics*).

a) *Dimensions* provides access to approximately 89 million scientific publications (of which 12.4% are open access), connected by nearly 900 million citations, as well as 20 million academic author profiles.
*https://app.dimensions.ai*
b) *Dimensions plus* gives access to the complete coverage of the database (adding patents, clinical trials, grants and policy documents), cyphered in 124 million documents, and their connections. Additionally it allows the search of new entities (organizations and financing agents). Finally, it provides access to the API (*Application Programming Interface*) (**Mori** and **Taylor**, 2018).

c) *Dimensions analytics* also includes advanced analysis tools, such as the comparison between organizations or financing agents, the generation of advanced reports, as well as the possibility of integrating custom implementations.

The production of the database started with the creation of a metadata backbone, from a wide range of sources (both open and licensed), among which *PubMed*, *PubMed central*, *Arxiv* and, most especially, *Crossref* stand out. After obtaining the metadata, the system proceeded to the full-text analysis of the documents, completed for about 55.5% (approximately 50 million) of the total indexed documents.

In the case of documents published under license, *Dimensions* collaborated with more than 100 academic publishers to index full text documents and improve the user experience in search and discovery tasks, in a similar way to the one operated by *Google scholar*. Although a master list of the publishers is not offered (only those that contribute more documents: *Elsevier*, *Springer Nature*, *Wiley*, *Ieee*, *Taylor & Francis*, *Sage publications*, *Cambridge university press*, *Wolters kluwer*, *DeGruyter*, *Oxford university press*, *Royal society of chemistry*), these can be intuited when performing different searches on the platform. Whatever the case may be, the coverage is over 50,000 journals (**Bode** et al, 2018). In the free version of *Dimensions*, the included journals are delimited by four sources: *Doaj*, *ERA list*, *Norwegian register for scientific journals, series and publishers* and *Pubmed*. On the other hand, *Web of science master journal list* is in beta mode to be incorporated.

The subject classification has been carried out at the document level instead of at the journal level (the usual procedure followed by databases such as *Web of science* or *Scopus*). For this purpose, existing classification systems (in this case the *Australian and New Zealand standard research classification*, ANZSRC) as well as artificial intelligence and machine learning techniques have been used to classify not only the articles but also the rest of the documents included (clinical trials, awarded grants and patents). In this way, 22 large areas (coded with two digits) and 154 subareas (coded with four digits) have been established. For example, the 'Library and Information Studies' sub-area corresponds to code 0807, within the 'Information and Computing Sciences' area (code 08).
*http://www.abs.gov.au/AUSSTATS/abs@.nsf/DetailsPage/1297.02008?OpenDocument*

Given the short time elapsed since its launch, there is currently limited scientific literature that has addressed the study of this database. It is necessary to highlight the descriptive note by **Orduña-Malea** and **Delgado López-Cózar** (2018), and the pioneering empirical study of **Thelwall** (2018), who after analysing a sample of 10,000 articles in the area of 'Food science' in the period 2008-2018, finds strong correlations between *Dimensions* and *Scopus*.

With the purpose of verifying and contrasting these preliminary results as well as complementing them with new empirical data, this work proposes, on the one hand, to describe the main characteristics of *Dimensions* as a search engine (functionalities, strengths, and limitations) and, on the other hand, to perform an analysis of its coverage at three levels (journals, documents and authors).

## 2. Objectives

The main objective of this work is to make a general description of the free version of *Dimensions*. To this end, the following specific objectives are proposed:

a) To describe the functionalities and search features of *Dimensions*, identifying their main strengths and weaknesses.
b) To analyse the coverage of *Dimensions* in order to determine whether the metric indicators offered have an order of magnitude significant enough to be used in bibliometric studies.

## 3. Method

The description of the tool has been performed directly from the free version of *Dimensions,* where all the functionalities of search, filters, ordering of results, descriptive information, and available reports have been tested and experimented. A special attention has been paid to the accuracy of the system, possible errors (due to information silence or noise), thematic assignment to the documents, and the accuracy in the information of the bibliographic records.

Regarding the quantitative analysis, a study has been carried out at three levels (journal, document and author), which is detailed below:

a) *Journal level*: based on a set of journals from the same discipline (top 20 'Library and information science' journals according to *Google scholar metrics*, as of the latest available edition of 2017), we have proceeded to calculate the h5-index (2012-2016) for these journals in *Google scholar metrics* (GSM) as well as in *Scopus* and *Dimensions*). https://scholar.google.com/citations?view_op=top_venues&hl=en&vq=soc_libraryinformationscience

b) *Document level*: considering one academic journal specialized in Bibliometrics (*Journal of informetrics*) during a recent period (2013, 2014 and 2015), the number of indexed documents, the number of citations received, as well as the *Altmetric attention score* received by each document, have been gathered and calculated. In order to compare results, we have obtained the number of citations received for those same documents from *Scopus*.

c) *Author level*: the number of citations received and the h-index have been calculated for all the authors who won the *Derek de Solla Price* prize (a total of 28 authors to date). The data has been captured both from *Google scholar citations* (directly from the authors' public academic profiles) and from *Scopus* and *Dimensions*. In these two last cases, data have been calculated considering the different variants of the existing name. http://www.issi-society.org

All data were captured and analysed in February 2018.

## 4. Results

### 4.1. *General description of Dimensions*

**a) Basic operation**

The *Dimensions* starting screen can be divided into four clearly distinguishable zones: 1) the search box; 2) the results page; 3) the filters; and 4) the analytical reports (Figure 1).

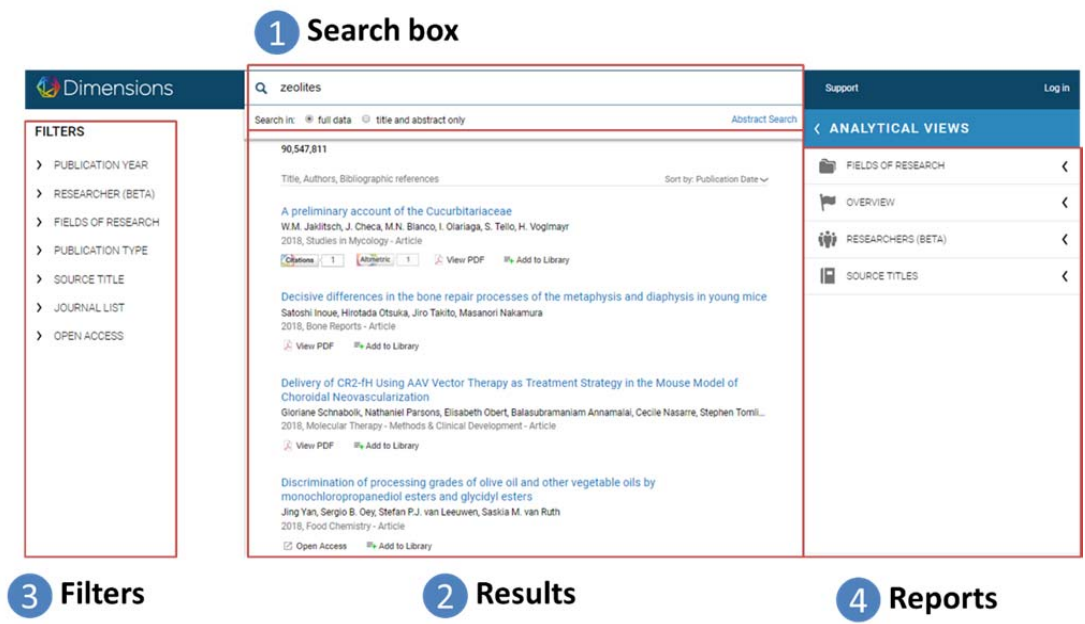

**Figure 1. Information zones in the *Dimensions* starting screen**
Source: *https://www.dimensions.ai*

One of the first sensations that the user perceives of the system is complete transparency in terms of data. The product offers precise statistics of the amount of documents (totals, per year, per author, per source, per type, etc.) available, both relative to the database in general and to those corresponding to a query in particular.

The search box allows two types of search (by keyword or by abstract). The keyword search can also be performed on the full text of the documents (available for 55.5% of the records) or just on the abstract and title.

Once the search is completed, the user can apply any of the existing filters: year of publication, researcher, field of knowledge, type of publication (article, chapter, proceeding, monograph, preprint), source of the publication, list of journals (*Norwegian register*, ERA 2015, *PubMed*, *Doaj*) and finally by documents in open access. The filters corresponding to author, source and field of knowledge, in addition to suggesting filtering by those entities with a higher frequency of appearance among the results of the search, allow the execution of an advanced search from an independent search box.

Although the area of analytical reports might sound repetitive at first (and partly is), its main characteristic is to offer a set of reports with value added information, relating to the results of the query performed. To do this, the user must click directly on each available view (Overview, Fields of research, Researchers and Source titles). At that time, the reporting area is displayed, hiding the results zone.

The Overview sight offers the time evolution related to the number of articles published, the number of total citations received, the percentage of documents cited (and not cited), the number of citations per publication, the average value of the RCR (Relative Citation Ratio) and the average value of the FCR (Field Citation Ratio). The 'Fields of research', 'Researchers' and 'Source titles' views are very similar, offering the number of articles according to each discipline, author and publication respectively, together with the values of RCR and FCR at the level of these entities (always relative to the query completed).

*Dimensions* database offers additional access to profiles at author and journal level (to access them, simply click on the name of the author or journal in the results offered by the different analytical views). In the case of the author, her/his institutional affiliation and the total number of citations received are showcased. As regards journals, the SNIP (Source normalized impact per paper) and the SJR (Scimago journal & country rank) are displayed.

Finally, the area of results provides the list of records that respond to the search queried. The user can sort these results according to various parameters (relevance, date of publication, RCR, citations received, and *Altmetric attention score*). Each bibliographic record includes the title, authors, source and date of publication. Additionally, two badges are offered, one corresponds to the number of citations and the other to the Altmetric score. If the user places the mouse over the badges, they will show the disaggregated data for both citations (total citations, recent citations, FCR, and RCR) and altmetrics (*Reads* in *Mendeley*, *Tweets* in *Twitter*, mentions in *Wikipedia* pages, etc.). When clicking directly on the badges, the system redirects the user to a personalized page of metrics at author level. While the altmetric badge redirects the user to the personalized page available by *altmetrics.com*, the citation badge redirects the user to a similar page, although within the *Dimensions* web domain, being therefore an original development within this product.

The fact that the system works even without performing a query allows the user to obtain interesting facts. For example, the article with the highest number of citations received ("Cleavage of structural proteins during the assembly of the head of bacteriophage T4", by Laemmli, with 176,406 citations). A fact that by the way contradicts the data that indicate that the most cited article in history, both by WoS and Google Scholar, is "Protein measurement with the Folin phenol reagent", by Lowry et al. (Martín-Martín et al., 2015), which appears in the third position of the Dimensions ranking (93,316 citations). However, data should be taken cautiously since as on April 2sd, Lowry's article climbs to the first position (203,490 citations). This huge change in the number of citations (110,174 citations in just a month), which can hardly been attributable to a natural increase in the coverage of the database, shows similarities with the anomalies detected by the literature for this same article in Google Scholar (Martín-Martín et al., 2016). A simple comparison of the 10 most cited articles in Dimensions with respect to Web of Science (Table 1) shows the differences not only in the number of citations captured but in the order in which they place the articles according to citations received in both databases. This reminds us how the nature and coverage of a database necessarily conditions its search results and its metrics.

**Table 1. Top 10 documents with a higher number of citations received in *Dimensions* and the *Web of Science* (*Alldatabases*)**

| AUTHOR | TITLE | CITATIONS | |
|---|---|---|---|
| | | *DIMENSIONS* | *WOS* |
| UK Laemmli | Cleavage of Structural Proteins during the Assembly of the Head of Bacteriophage T4 | 176,406 | 245,702 |
| MM Bradford | A rapid and sensitive method for the quantitation of microgram quantities of protein utilizing the principle of protein-dye binding | 153,340 | 203,260 |
| O H Lowry et al | Protein measurement with the Folin phenol reagent | 93,316 | 336,943 |
| KJ. Livak, TD. Schmittgen | Analysis of Relative Gene Expression Data Using Real-Time Quantitative PCR and the 2−ΔΔCT Method | 58,438 | 61,347 |
| AD. Becke | Density-functional thermochemistry. III. The role of exact exchange | 54,906 | 66,555 |
| J P. Perdew, K Burke, M Ernzerhof | Generalized Gradient Approximation Made Simple | 52,419 | 66,328 |
| G.M. Sheldrick | A short history of SHELX | 51,597 | 64,887 |
| MF. Folstein, Susan E. Folstein, Paul R. McHugh | "Mini-mental state" A practical method for grading the cognitive state of patients for the clinician | 49,962 | 44,995 |
| F. Sanger, S. Nicklen, A. R. Coulson | DNA sequencing with chain-terminating inhibitors | 49,770 | 66,929 |
| CLee, W Yang, RG. Parr | Development of the Colle-Salvetti correlation-energy formula into a functional of the electron density | 49,177 | 62,931 |

Nevertheless, one of the most important novelties of *Dimensions* is the possibility of knowing which documents with the highest altmetric impact actually are, through the data supplied by *Altmetric.com*. Since this parameter is offered as a criterion for document sorting, it can be applied to any search and aggregation unit (author, journal, field, discipline). Thus, we can ascertain that the document with the higher *Altmetric attention score* corresponds to "How diversity Works", by Katherine W. Phillips, with a score of 11,672 (with 14,229 mentions in *Twitter*), followed surprisingly by two articles signed by Barack Obama (44th President of the United States from 2009 to 2017). A simple comparison between Tables 1 and 2 allows us to realize the different nature of the documents as measured by the number of citations and the "altmetric attention". Measures that have nothing to do with each other.

**Table 2. Top 10 documents with the higher number of *Altmetric attention score* in *Dimensions***

| AUTHOR | TITLE | ALTMETRIC SCORE |
|---|---|---|
| KW Phillips | How Diversity Works | 11,672 |
| B Obama | United States Health Care Reform: Progress to Date and Next Steps | 8,362 |
| B Obama | The irreversible momentum of clean energy | 7,813 |
| R Nuzzo | Scientific method: Statistical errors | 7,638 |
| WJ. Ripple et al | World Scientists' Warning to Humanity: A Second Notice | 7,088 |
| R Cowen | Simulations back up theory that Universe is a hologram | 6,983 |
| M Dehghan et al | Associations of fats and carbohydrate intake with cardiovascular disease and mortality in 18 countries from five continents (PURE): a prospective cohort study | 6,837 |
| R Van Noorden | Nature promotes read-only sharing by subscribers | 6,548 |
| T Mizuno, H Kubo | Overview of active cesium contamination of freshwater fish in Fukushima and Eastern Japan | 6,560 |
| R Van Noorden | Online collaboration: Scientists and the social network | 6,532 |

The fact that *Dimensions* offers a subject categorization at the article level, apart from being an important technological novelty, opens the door to multilevel thematic studies both at the journal and at the author level. In Figure 2, we can observe the results obtained for a multidisciplinary journal (*PLos one*) and for a specialized journal (*Journal of informetrics*). In a fast way the user can know, for example, that within *PLoS one*, the articles of 'Genetics' (32,203), 'Biochemistry and Cell biology' (31,218) and 'Clinical sciences' (24,395) constitute a fundamental part of the journal. This analysis is even more interesting when applied to a specialized journal. In this way, 'Applied economics' (133), 'Information systems' (94) and 'Psychology' (71) are the areas with the highest representation in the *Journal of informetrics*.

*Dimensions* database provides the FCR and RCR data linked to each field of knowledge within the documents published by each journal. In this manner, we can find out that in *PLoS one*, the correlation between the number of documents assigned to one area of knowledge and the corresponding FCR presents weak and negative values (Rs = -0.34), being 'Agriculture, land and farm management' (FCR: 14.64 ) and 'Literary studies' (FCR: 13.33) the areas with the highest FCR within that journal. In the case of *Journal of informetrics*, the correlation between the number of documents assigned to an area of knowledge and the FCR is similar, although of positive value (Rs = 0.37), being 'Literary studies' (58.87) and 'Philosophy' (48.9) the areas with greater FCR in the journal.

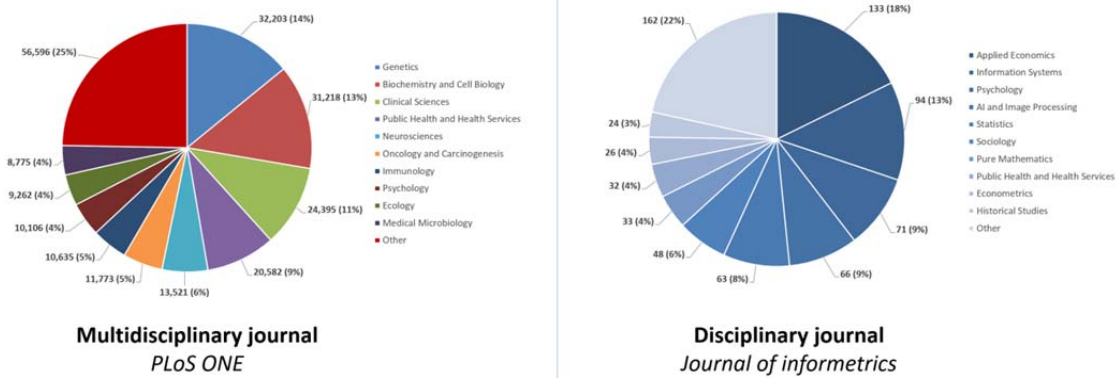

**Figure 2. Journal multilevel analysis in *Dimensions***
**Left: multidisciplinar journal (*PLoS One*)**
**Right: disciplinary journal (*Journal of informetrics*)**
Source*: https://www.dimensions.ai*

Similar data but this time at author-level is offered in the Figure 3, where we can know the productivity of authors in different areas of knowledge, especially useful for those authors who have cultivated different fields with remarkable performance.

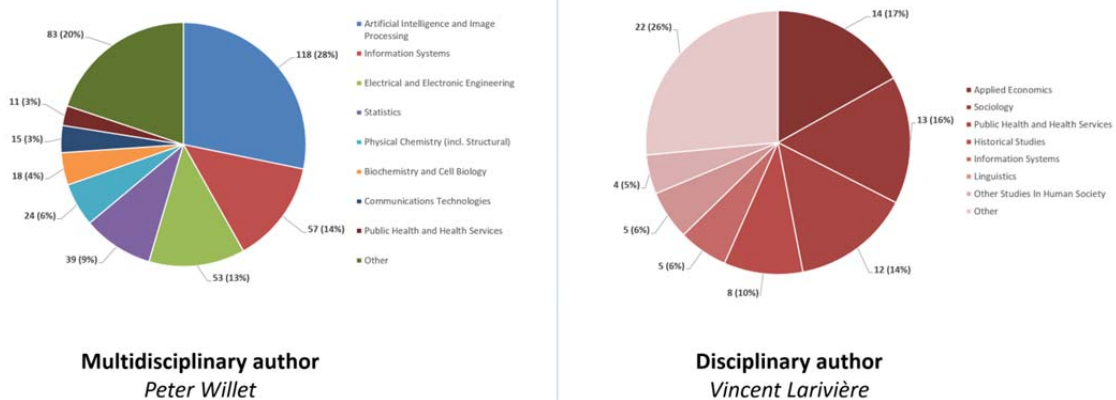

**Figure 3. Author multilevel analysis in *Dimensions***
**Left: multidisciplinary author (*Peter Willet*)**
**Right: disciplinary author (*Vincent Larivière*)**
Source*: https://www.dimensions.ai*

Otherwise, by clicking on the hyperlink of the title of each result, the user can access the full bibliographic description of each document, which includes not only the descriptive information, but also the abstract, document references, document citations, supporting grants, patent citations, and linked clinical trials. Additionally, the citation badge (total citations and recent citations [last two years]) and the altmetric badge are too available for each document. Users can also embed the badges on their websites or reuse them in other applications.
*https://badge.dimensions.ai*

*Dimensions* allows users to store any document offered in the results list of any query in a personal library, similar to how *My library* operates in *Google scholar*. The difference lies, in this case, in that the "Add to Library" button (available both on the results page and on the descriptive page of each document) allows the user to save and organize their favorite references through the *ReadCube* cloud library service. Finally, users can register in the system and then save search query strategies.
*https://scholar.google.com/intl/en/scholar/help.html#library*
*https://www.readcube.com*

**b) Strengths and weaknesses**

During the analysis, a series of limitations have been identified, some of them significant, that deserve to be highlighted.

The search by countries and institutions is disabled in the free version. Even though these options are not critical for an average user (or for anyone who does not intend to perform quantitative analyses), they do imply a lack of benefits, especially when it comes to the search for university general output. Whilst it is necessary to recognise that certain features are left to the paid versions, the lack of institutional filtering is a limitation in the free version of *Dimensions*.

With regard to the design of advanced queries, the user must perform a search and then add the relevant filters. However, you cannot implement all the filters at once; you must go one by one. For example, the user cannot select two years of publication and an area of knowledge at the same time; you must mark the years first, execute the action, and then filter by area. Otherwise, if we wished to eliminate one of the two selected years, we cannot do it directly, we must eliminate the two years in block and, later, filter for the desired year (only one complete filter block can be deleted). These issues, even when minor, suppose a slowdown in continuous search processes guided by serendipity and discovery.

The search by authors involves another important shortcoming. First, there is a large amount (to be determined) of duplicate authors. Therefore, the authority control has not worked completely well. When the user activates the secondary search box of authors and manually enters the name of the author, the system suggests the names matching the written string of characters. This is the moment in which duplicates are noticed (in many cases the variant of the name is identical) and sensitivity to diacritics is observed. As an illustrative example, the number of variants identified for **Eugene Garfield** is shown in Figure 4. Moreover, depending on how the name of the author is written ('name surname'; 'surname, name'; 'surname, initial name', etc.), the system may or may not suggest the desired names even if they are already included in the database. For example, if the term "Cronin" is written, the system suggests up to nine names (none of them **Blaise Cronin**). However, if you directly type "Blaise Cronin", the system detects three identical variants of the same name. *Dimensions* has announced that the integration of *Orcid* is planned to solve part of the problem.

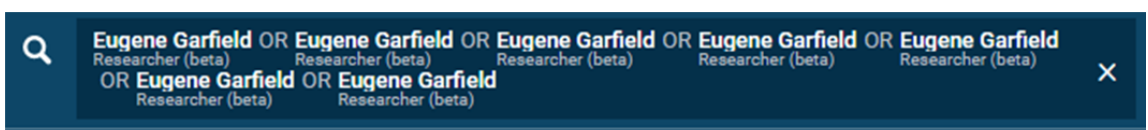

**Figure 4. Inconsistencies in the identification of authors in Dimensions.**
The author search engine identifies exactly identical records (same first name and same last name, written the same) but that are independent, and each one of them returns different results, even being the same author. In this case of Figure 4 they are all "identical variants" of Eugene Garfield.
Source: *Dimensions*

Inconsistencies in the indexing of journal articles, which are made cover-to-cover, are also observed. As the process is carried out automatically, the system incorporates as articles items such as "List of reviewers" or "Editorial Board", which may be inflating the total number of items indexed by the system.

As regards the subject classification, it presents important limitations. On the one hand, many articles (in a proportion to be determined) appear uncategorized. On the other hand, it seems to work mostly for articles in English. For example, if a user tries to perform the analysis for the journal "Profesional de la información", the system returns the following message: "There are no fields of research matching your search".

However, the most serious aspect seems to fall on the precision of the automatic categorization carried out (see Figures 2 and 3). For example, the most cited article published in *Journal of informetrics* according to *Dimensions* corresponds to "h-Index: A review focused on its variants, computation and standardization for different scientific fields", a bibliographic review that has been classified within "Literary studies" (category 2005). A manual review of the 151 articles published by *PLoS one* and classified within "Historical studies" allows us to contrast that the results contain abundant classification errors (for example, the *Plos One*'s most cited article within that category corresponds to a bibliometric study, "Why Has the Number of Scientific Retractions Increased?"). The same impression is obtained when reviewing the categories "Applied economics" or "Sociology", within this same journal. This inclines us to invalidate the results previously offered in figures 2 and 3. These "anecdotal evidences" make us suspect on the reliability and general validity of the subject classification used. The importance of this issue goes beyond a correct or deficient classification of a document in the database, since it affects the goodness of the bibliometric indicators. Since these indicators (mainly FCR and RCR) are normalized according to the field of knowledge and discipline, the incorrect composition of fields and disciplines invalidates the indicators obtained. Therefore, even when more empirical studies are needed to test the accuracy of this classification, users should be warned about its use.

Another important problem, already pointed out by **Andrew Gray**, is the subject categorization of the monographs, which apparently only takes place from the abstract, which generates a series of inconsistencies at present.
*https://twitter.com/generalising/status/953237327635189760*

Finally, another type of unexpected errors are those related to certain bibliographic data. For example, Figure 5 shows how the year of publication of an article has been modified. Whilst it is well described in the original source (date of publication equal to 2010), it exhibits an error in *Crossref* (date of publication equal to 2009), from where the error has probably been inherited.

**Figure 5. Inconsistencies in the date of publication en *Dimensions***
Source: *Dimensions*

However, the innumerable advantages of the product must also be highlighted, among which the following stand out:

In the first place, and as already indicated, the transparency of the system spotlights, providing all kinds of updated statistics and data. Likewise, there is an implication of the developers in listening to the users' experience in an open and participatory way.

In the second place, the fact that all filters and analytical views are activated automatically after each search allows the user the possibility to get a very quick idea of certain properties of the data. For example, a simple full text search for "zeolites" quickly indicates that 383,925 documents are located (there is consequently abundant literature). The main field of research is "Physical Chemistry" with connections to "Materials engineering", "Chemical engineering" and "Inorganic engineering". There is a productivity peak between 2013 and 2014, and Dr. Avelino Corma (with 897 documents published) is the most productive in the world on the subject. The most published journals on the subject are "Applied catalysis A (General)", "Microporous and mesoporous materials" and "Journal of catalysis". In addition, bibliometric data can be obtained both for general data (in Overview) and for relative to authors and journals (in their corresponding views) quickly and easily.

On the other hand, the open access filter is implemented at the article level (and not the journal), which allows users to access hybrid journals (which only publish a part of their articles in open access) of a simple and integrated with the rest of search functionality, aspect already highlighted on *Twitter* by **Andrew Gray**.

## 4.2. Coverage analysis

As previously mentioned, the system is transparent. For this reason, it is practically trivial to know the coverage and evolution of the database. *Dimensions* covers the scientific literature from 1665 to the present. In Figure 6 we can see its comparative size with respect to the *Web of Science Core Collection* (WoScc) and *Scopus* (from 2000 to 2016), where it shows a higher annual coverage. The data of *Google Scholar* present great inconsistencies in the searches per year (especially from 2012 onwards) that have prevented its calculation by previously applied methods (**Orduna-Malea** et al, 2015). For that reason, it is shown for illustrative purposes only.

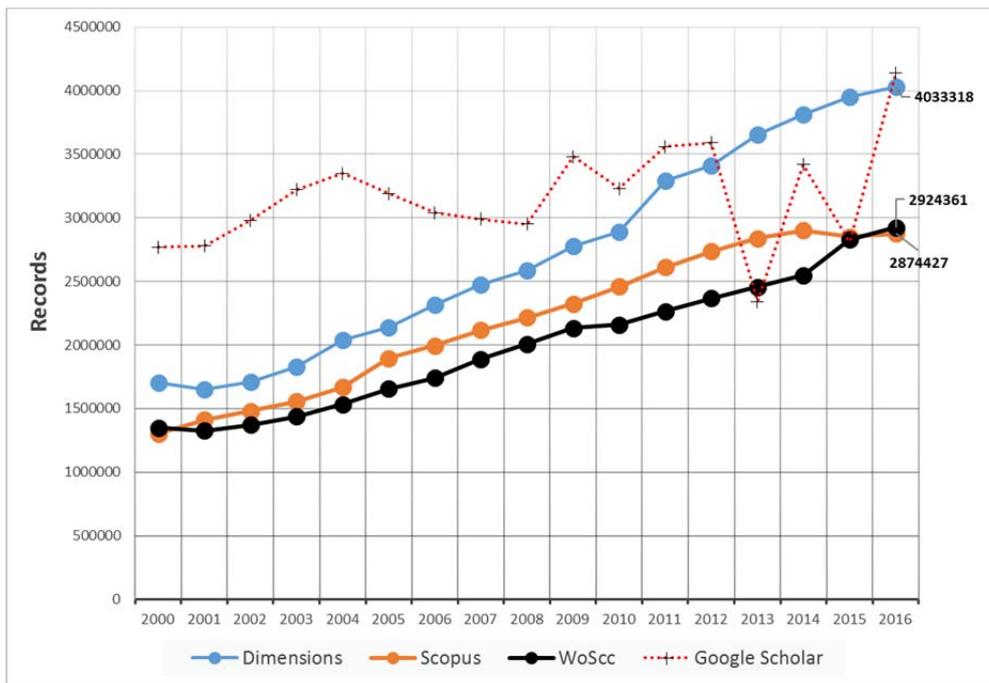

**Figure 6. Evolution of the number of records (*Dimensions*, *Scopus*, *Web of science core collection*)**
Note: data from Google Scholar are gathered through a null query filtered by year, and removing "citations" and "patents".

### a) Journal level

The h5-index (2012-2016) corresponding to 17 journals of 'Library Science and Documentation' according to *Dimensions* is offered in Table 3. These results are compared with those corresponding to *Scopus* and *Google scholar metrics*.

**Table 3. H5-Index of 'Library Science and Documentation' journals (*Google scholar metrics*, *Scopus* y *Dimensions*)**

| JOURNALS | H5-INDEX (2012-2016) | | |
|---|---|---|---|
| | *GSM* | *SCOPUS* | *DIMENSIONS* |
| *JASIST* | 53 | 43 | 42 |
| *Scientometrics* | 49 | 39 | 37 |
| *Journal of informetrics* | 36 | 31 | 30 |
| *Online information review* | 26 | 20 | 19 |
| *College & research libraries* | 25 | 19 | 15 |
| *Library & information science research* | 25 | 17 | 20 |
| *The journal of academic librarianship* | 25 | 19 | 16 |
| *Journal of documentation* | 22 | 16 | 18 |
| *Journal of information science* | 22 | 16 | 16 |
| *Library hi tech* | 20 | 15 | 13 |
| *Reference services review* | 20 | 13 | 11 |
| *The electronic library* | 20 | 13 | 12 |
| *Portal: libraries and the academy* | 20 | 15 | 12 |
| *Aslib journal of information management* | 20 | 13 | 12 |
| *Journal of the medical library association* | 19 | 15 | 15 |
| *Journal of librarianship and information science* | 19 | 12 | 10 |
| *New library world* | 18 | 13 | 11 |

Note 1: In the case of *JASIST* and *ASLIB*, the change in the title of the journal has been taken into account.
Note 2: GSM data correspond to June 2017; *Dimensions* and *Scopus* data correspond to February 2018.
Note 3: *ArXiv Digital Libraries* (repository), *International ACM/IEEE Joint Conference on Digital Libraries* (conference proceedings) y *Proceedings of the Association for Information Science and Technology* (no included in Scopus) have been excluded.

As can be seen, GSM shows higher values, even taking into account that the values correspond to June 2017, so it is likely that they will be even higher today. *Scopus* and *Dimensions* offer very similar values, *Scopus* being slightly higher in most cases, although some exceptions are detected ('Library & information science research' and 'Journal of documentation'). Overall, the h5-index values (2012-2016) obtained in *Dimensions* strongly correlate in a statistically significant way both with *Scopus* (Rs = 0.94) and with GSM (Rs = 0.90).

**b) Document level**

The coverage of the *Journal of informetrics* in *Dimensions* for the years 2013, 2014 and 2015 amounts, after eliminating various non-relevant documents (a total of 15, including 'List of reviewers' and 'Editorial boards'), to a total of 276 documents (in *Scopus* 279 are obtained), of which only 41.7% receive Altmetric attention scores (Table 4).

With regard to the number of citations accumulated (summation of citations of each article), *Scopus* is slightly higher except in 2015, where *Dimensions* exceeds *Scopus*. This result could corroborate, together with the size data previously seen in Figure 6, that *Dimensions* offers a higher coverage of the most recent literature.

**Table 4. Presence of the *Journal of informetrics* in *Dimensions* (2013-2015)**

| Metrics | 2013 | 2014 | 2015 | TOTAL |
|---|---|---|---|---|
| Nº of documents | 103 | 89 | 84 | **276** |
| Nº of documents with *Altmetric score* | 41 | 31 | 43 | **115** |
| Accumulated *Altmetric scores* | 366 | 129 | 1051 | **1,546** |
| Accumulated citations (*Dimensions*) | 1,392 | 857 | 655 | **2,904** |
| Accumulated citations (*Scopus*) | 1,416 | 901 | 648 | **2,965** |

Regarding the *Altmetric attention scores*, no clear trends are evident (three years do not allow a very broad follow-up). In any case, the values obtained in 2015 stand out, year in which 51.2% of the articles of the journal indexed in *Dimensions* have an *Altmetric score*. The accumulated values are affected by the special case of the article "Attention decay in science", which obtains a score equal to 769.

On the other hand, a very strong and significant correlation was observed between the citations in *Scopus* and in *Dimensions* (Rs = 0.96; α = 0.1) and practically non-existent between citations and the *Altmetric score* in *Dimensions* (Rs = 0.28; α = 0.1). These last results are aligned with those previously obtained in the scientific literature (**Costas, Zahedi, Wouters**, 2015, **Thelwall** et al, 2013, **Thelwall**, 2018). These results are illustrated graphically in the Figure 7, where you can see how documents with high *Altmetric attention scores* are not necessarily the most cited and vice versa.

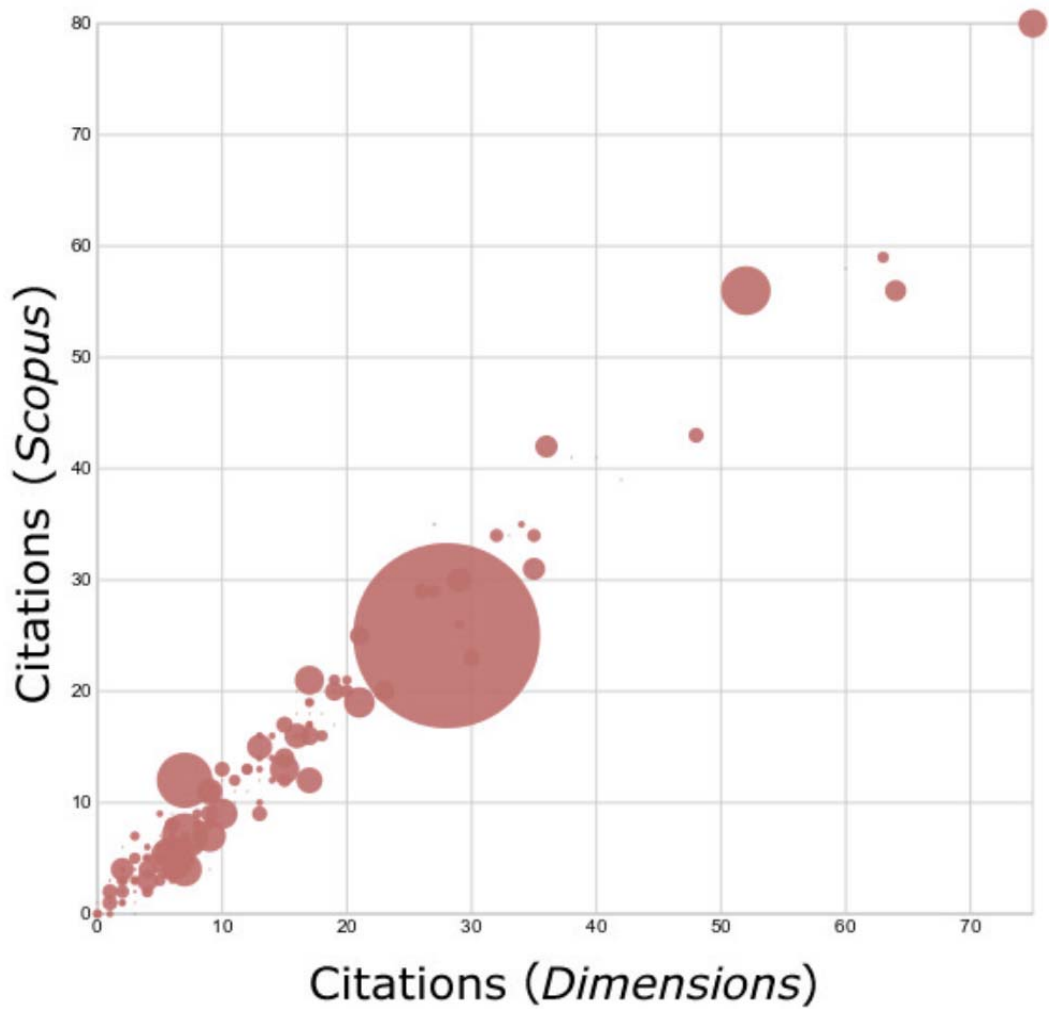

**Figure 7. Scatterplot of the number of citations in *Dimensions* and *Scopus***
Source: *Dimensions*
Node size: value equivalent to the *Altmetric attention score*.

**c) Author level**

Finally, the coverage results at author-level is displayed in the Table 5. In this case, the number of citations received by each author and their h-index are shown according to *Dimensions*. Additionally, these same data are offered according to *Scopus* and *Google scholar citations*.

In the specific case of *Dimensions*, the number of variants per author has been included. For example, up to seven variants have been detected in the case of **Garfield** and six in the case of **Egghe**. This circumstance has made it necessary to dispense with the information provided by the *Dimensions'* author profiles and to proceed calculating this data considering all the documents corresponding to all the variants manually.

**Table 5. Citations received and h-index for the authors awarded with the *Derek de Solla Price* (*Scopus*, *Google Scholar Metrics* and *Dimensions*)**

| Author | *Scopus* | | *GSC* | | *Dimensions* | | |
|---|---|---|---|---|---|---|---|
| | **Citations** | **H-Index** | **Citations** | **H-index** | **Citations** | **H-index** | **Duplicates** |
| Garfield, Eugene | 7,593 | 33 | 29,881 | 62 | 1,240 | 13 | 7 |
| Moravcsik, Michael J | 1,599 | 20 | 4,963 | 30 | 184 | 8 | 2 |
| Braun, Tibor | 4,638 | 33 | 9,035 | 47 | 1,237 | 18 | 1 |
| Nalimov, Vasily V | 32 | 2 | 10,503 | 36 | N/A | N/A | N/A |
| Small, Henry | 4,260 | 28 | 9,739 | 38 | 855 | 16 | 4 |
| Narin, Francis | 5,305 | 32 | 14,527 | 48 | 2,674 | 22 | 3 |
| Brookes, Bertram C | 843 | 16 | 3,681 | 25 | 322 | 7 | 1 |
| Vlachy, Jan | 603 | 11 | 1,069 | 15 | 116 | 4 | 4 |
| Schubert, András | 5,463 | 35 | 11,571 | 52 | 3,235 | 25 | 3 |
| Van Raan, Anthony FJ | 5,556 | 41 | 12,359 | 58 | 4,012 | 36 | 3 |
| Merton, Robert K | 4,122 | 17 | 158,497 | 114 | N/A | N/A | N/A |
| Irvine, John | 908 | 17 | 3,406 | 26 | N/A | N/A | N/A |
| Martin, Ben | 4,303 | 30 | 13,261 | 48 | 2,185 | 15 | 1 |
| Griffith, Belver C | 1,628 | 15 | 8,089 | 29 | 1,306 | 12 | 1 |
| Glänzel, Wolfgang | 7,280 | 46 | 15,192 | 65 | 3,509 | 33 | 2 |
| Moed, Henk F | 4,707 | 36 | 10,942 | 53 | 3,183 | 31 | 1 |
| Rousseau, Ronald | 4,782 | 33 | 11,807 | 50 | 1,935 | 22 | 1 |
| Egghe, Leo | 4,782 | 33 | | | 2,146 | 17 | 6 |
| Leydesdorff, Loet | 7,855 | 58 | 39,966 | 89 | 9,712 | 50 | 1 |
| Ingwersen, Peter | 2,742 | 23 | 10,244 | 37 | 784 | 16 | 3 |
| White, Howard D | 9,307 | 42 | 6,984 | 32 | 2,451 | 26 | 3 |
| McCain, Katherine W | 2,226 | 22 | 5,881 | 31 | 1,602 | 18 | 1 |
| Vinkler, Péter | 1,496 | 23 | N/A | N/A | 813 | 17 | 2 |
| Zitt, Michel | 1,053 | 19 | N/A | N/A | 700 | 14 | 1 |
| Persson, Olle | 1,887 | 19 | 4,570 | 27 | 1,178 | 15 | 2 |
| Cronin, Blaise | 3,327 | 28 | 11,401 | 52 | 1,949 | 24 | 3 |
| Thelwall, Mike | 10,205 | 53 | 22,480 | 77 | 7,560 | 45 | 1 |
| Bar-Ilan, J | 3,016 | 30 | 6,683 | 43 | 1,318 | 20 | 1 |

GSM: *Google Scholar citations*
N/A: *No data available*

The data offered in Table 5 shows how *Google scholar metrics* is the product with the greatest coverage, followed by *Scopus* and finally *Dimensions*. Despite this, we find unexpected exceptions. For example, in the case of **Leydesdorff**, whereas 7,855 citations are detected in Scopus, we find 9,712 in *Dimensions* (11,573 as of April 3rd; which confirms a huge coverage growth in just few weeks). This may reflect the existence of inconsistencies in the authority control system and the need to filter through document to document to ensure the non-inclusion of articles that did not correspond to the author.

In any case, the correlations obtained between the metrics of the different databases are equally significant and very strong, especially between *Dimensions* and *Scopus*. In terms of the number of citations received per author, the correlation between *Dimensions* and *Scopus* amounts to 0.81, while the correlation of *Dimensions* with *Google scholar metrics* is somewhat lower (Rs = 0.74). Analogously, at the h-index level, the correlation between *Dimensions* and *Scopus* is 0.88, while between *Dimensions* and *Google scholar metrics* is 0.79. In these results, it should be taken into account that the profiles in *Scopus* and *Dimensions* are automatic (without the authors' intervention) while in *Google scholar citations* they depend as much on the creation (on the part of the author or a third party) as on their optimal management and ethics (excluding documents not authored by the authors).

## 5. Final remarks

This work is a first exploratory evaluation of the free version of *Dimensions*. It has been carried out on the one hand a description of its operation (highlighting some of its strengths and weaknesses) and, on the other hand, an analysis of its coverage in order to know the magnitude of certain bibliometric indicators (analysing a sample of journals, documents and authors).

Regarding the database system operation, in spite of certain inconsistencies and errors (especially those related to the authority control and the subject classification), it is concluded that *Dimensions* is a product with potential. It combines certain functionalities of the classic bibliographic databases (availability of metrics, search filters, and sort results functionality) with some of the characteristics of *Google scholar* (greater coverage, simple search box, fast and precise, and free). In any case, the analysis of *Dimensions plus* and *Dimensions analytics* should give a more complete drawing by adding not only a greater coverage (and semantic linkage between documents), but a set of built-in analysis tools, giving meaning to the product philosophy based on generating evidences and connecting the different stages of scientific activity.

Regarding the coverage analysis, the results indicate that the annual growth (number of records per year) is currently higher in *Dimensions* than in *Scopus* or *Web of science core collection*, also exceeding them in the number of total records. However, at the level of citations received, *Dimensions* is slightly below *Scopus*, although the correlation of the different metrics at different levels (author, document, and journal) is very high. These data are aligned with results obtained by **Thelwall** (2018), although with completely different samples. This circumstance is of special relevance, given the free nature of this tool. In any case, it should be taken into account that the samples analysed in this work are very small and biased to a very specific field of knowledge (Library & Information science), so that different results could be obtained in other disciplines.